\documentclass[preprint,showpacs,onecolumn,nofootinbib]{revtex4}
\pdfoutput=1
\usepackage[colorlinks=true,linkcolor=blue,urlcolor=blue,filecolor=black,citecolor=red,pdfstartview=FitV,pdftitle={},pdfsubject={},pdfkeywords={},pdfpagemode=None,bookmarksopen=true]{hyperref}
\usepackage{graphicx}%Include figure files
\usepackage{amsfonts}
\usepackage{amssymb,ulem}
\usepackage{color}%
\usepackage{dcolumn}% Align table columns on decimal pointl
\usepackage{amsmath}
\usepackage{epsfig}
\usepackage{graphics}
\usepackage[active]{srcltx}
\usepackage{epstopdf}
\usepackage{shuffle}
\usepackage[utf8]{inputenc}
\usepackage{pifont}
\usepackage{appendix}

\newcommand{\bea}{\begin{eqnarray}}
\newcommand{\eea}{\end{eqnarray}}
\newcommand{\bean}{\begin{eqnarray*}}
\newcommand{\eean}{\end{eqnarray*}}
\newcommand{\nn}{\nonumber \\}

\def\eref#1{(\ref{#1})}

\def\Label#1{\label{#1}%
  \smash{\hbox to0pt{\raise1ex\hbox{\tiny[#1]}\hss}}}

\begin{document}

\baselineskip=0.6 cm
\title{Soft theorems based on differential operators from gravity to Yang-Mills and BAS}
\author{Fang-Stars Wei}
\email{MX120220339@stu.yzu.edu.cn}
\affiliation{Center for Gravitation and Cosmology, College of Physical Science and Technology, Yangzhou University, Yangzhou, 225009, China}

%\date{\today }
\begin{abstract}
\baselineskip=0.6 cm
This note study the soft behavior of Yang-Mills (YM) and bi-adjoint scalar (BAS) amplitudes at tree level, by using transmutation operators proposed by Cheung, Shen and Wen. By acting such transmutation operators to gravity amplitudes in the soft limit, we reproduce universal soft factors of YM amplitudes at the leading and sub-leading orders, and explain that the analogous universal soft behavior does not exist at the sub-sub-leading order. Subsequently, by acting the same operators on YM amplitudes, we obtain the universal soft factor of BAS amplitudes at the leading order. Furthermore, we find that a "weaker" version of universal soft behavior of BAS amplitudes holds at the sub-leading order, if we exclude the special $4$-point case.
\end{abstract}
%\keywords{aa}

%\pacs{ 04.20.Gz, 04.20.-q, 03.65.-w}
\maketitle

%\tableofcontents

%%%%%%%%%%%%%%%%%%%%%%%%%%%%%%%%%%%%%%
\section{Introduction}
\label{sec-intro}
%%%%%%%%%%%%%%%%%%%%%%%%%%%%%%

In recent years, the soft theorems for scattering amplitude became an active area of research.
The soft theorems state the universal behavior of scattering amplitudes when one or more external momenta tends to zero. More explicitly, one can introduce a parameter $\tau$ to rescale the external momentum carried by the external particle $q$ as $k_q \to \tau k_q$, and take the limit $\tau\to0$. In the above soft limit, the $n$-point tree level amplitude of certain theories, such as gravity (GR) and Yang-Mills (YM), factorizes into the universal soft factor and the $(n-1)$-point level amplitude. 

Historically, the soft theorems for photons and gravitons were first studied in \cite{Weinberg:1965nx,Gross:1968in}. The soft theorem for gravitons was extended to higher orders, motivated by the symmetry at infinity. The single massless particle states are defined on null infinite, and the corresponding symmetry group on the null infinite of asymptotic flat spacetime is known as the BMS group invariance \cite{Strominger:2013jfa}. Such BMS symmetry serves as the foundation of soft theorems for GR amplitudes, indicates the leading and sub-leading soft theorems for GR amplitudes \cite{He:2014laa,Bondi:1962px}. Further more, the sub-sub-leading universal soft behavior of GR amplitudes are also found at the tree level \cite{Cachazo:2014fwa}. These soft theorems were first verified by employing the well known BCFW on-shell recursion relation, which construct higher-point tree amplitudes from relative lower-point on-shell amplitudes \cite{Britto:2005fq,Britto:2004ap}. Subsequently, the sub-leading order and sub-sub-leading order soft theorems of GR have been extended to arbitrary dimensional spacetime by using the CHY \cite{Cachazo:2014xea,Cachazo:2013hca,Cachazo:2013iea} formula \cite{Schwab:2014xua,Zlotnikov:2014sva}.

The leading and sub-leading soft theorems for tree YM amplitudes have also been discovered \cite{Dixon:2008gr,Arkani-Hamed:2008owk,Casali:2014xpa}. Comparing with the situation of GR amplitudes, an interesting question is, if the tree YM amplitudes have further soft theorems at higher order, such like the sub-sub-leading soft theorem for GR. Whether or not the higher order soft theorem for YM amplitudes do not exist, we want to understand the reason.

In this paper, we study the soft behaviors of YM and bi-adjoint scalar (BAS) amplitudes at the tree level, by using the transmutation operators proposed by Cheung, Shen and Wen \cite{Cheung:2017ems}. They found that the tree amplitudes of certain theories can be transmuted into those of other theories, through differential operators. This unifying relations for tree level amplitudes has been generalized to $1$-loop Feynman integrands \cite{Zhou:2021kzv,Zhou:2022djx,Chen:2023bji}. At the tree level amplitudes, the unifying relations has also been generalized to the (A)dS spacetime \cite{Tao:2022nqc}. For example, applying the differential operators to GR amplitudes creates the YM amplitudes. The same operators also transmutes YM amplitudes to BAS ones. By using these operators, it is possible to connect soft theorems for GR, YM and BAS amplitudes together.
Then, assuming the validity of soft theorems for GR amplitudes which are supported by underlying BMS symmetry, one can derive the corresponding soft behaviors of YM and BAS amplitudes by using transmutation operators, and study the existence of universal soft theorems at higher order.

In section \ref{Chapter 2}, we briefly introduced the conclusions from the references, including the soft theorem of gravity and differential operators. 

In section \ref{Chapter 3}, starting from soft theorems of gravity, we reproduce the leading and sub-leading soft theorems of Yang-Mills theory through differential operators, and explain the reason why the universal soft behavior does not exist at the sub-sub-leading order.

In section \ref{Chapter 4}, we use the similar method to obtain the leading order soft theorem of BAS theory. We also find that although the universal soft behavior of BAS amplitudes does not exist at the sub-leading order, but the only exception is the $4$-point case. In other words, tree BAS amplitudes with at least $5$ external legs factorize into an universal operator and a lower-point amplitude, at the sub-leading order.

%%%%%%%%%%%%%%%%%%%%%%%%%%%%%%%%%%%%
\section{ Brief review for background}
\label{Chapter 2}
%%%%%%%%%%%%%%%%%%%%%%%%%%%%%%%%%%%%

In this section, we display three soft theorems for GR \cite{Cachazo:2014fwa,Schwab:2014xua,Zlotnikov:2014sva}, and some differential operators \cite{Cheung:2017ems} which will be used in subsequent sections. The scattering amplitudes addressed in this article pertain to the tree level amplitudes in arbitrary spacetime dimensions. Massless particles under consideration satisfy on-shell condition $k^{2}_{i}=m^{2}=0$. For non scalar particles, the transverse wave condition $k_{i} \cdot \epsilon_{i}=0$ is also satisfied.

%%%%%%%%%%%%%%%%%%%%%%%%%%%%%%%%%%%%%%
\subsection{Soft theorems of gravity}
\label{Chapter 2, Section 1}
%%%%%%%%%%%%%%%%%%%%%%%%%%%%%%%%%%%%%%

The soft theorems of GR amplitudes reads
\bea
{\cal A}_{\rm G}(\sigma_{no})&=&{\cal A}_{\rm G}^{(0)}(\sigma_{no})+{\cal A}_{\rm G}^{(1)}(\sigma_{no})+{\cal A}_{\rm G}^{(2)}(\sigma_{no})+O\left(\tau^{2}\right)\nn
&=&\left[{S}_{h}^{(0)}(q)+{S}_{h}^{(1)}(q)+{S}_{h}^{(2)}(q) \right] {\cal A}_{\rm G}(\sigma_{no}/q)+O\left(\tau^{2}\right)
\,,~~~~\label{soft theorem of gravitation}
\eea
where ${\cal A}_{\rm G}(\sigma_{no})$ represents the $n$-point tree level gravity amplitude and ${\cal A}_{\rm G}(\sigma_{no}/q)$ encodes the $(n-1)$-point tree level gravity amplitude obtained by removing the $q$-th graviton. Here and after, we use the notation $\sigma_{no}$ to denote unordered sets among $n$ external gravitons. The $q$-th graviton is the soft one, namely, $k_{q} \to \tau k_{q}$, $\tau \to 0$. The factorization behaviors ${\cal A}_{\rm G}^{(0)}(\sigma_{no})={S}_{h}^{(0)}(q) {\cal A}_{\rm G}(\sigma_{no}/q)$, ${\cal A}_{\rm G}^{(1)}(\sigma_{no})={S}_{h}^{(1)}(q) {\cal A}_{\rm G}(\sigma_{no}/q)$ and ${\cal A}_{\rm G}^{(2)}(\sigma_{no})={S}_{h}^{(2)}(q) {\cal A}_{\rm G}(\sigma_{no}/q)$ are called the soft theorems at leading, sub-leading and sub-sub-leading orders, respectively.

The corresponding universal soft factors are given as
\bea
{S}_{h}^{(0)}(q) \equiv {1 \over \tau} \sum_{j} {k_{j} \cdot {\varepsilon}_{q} \cdot k_{j} \over s_{jq}}={1 \over \tau} \sum_{j} {(\epsilon_{q} \cdot k_{j})(\epsilon_{q} \cdot k_{j}) \over s_{jq}}
\,,~~~~\label{soft factor of leading order of gravitation}
\eea
\bea
{S}_{h}^{(1)}(q) \equiv \sum_{j}{k_{j} \cdot {\varepsilon}_{q} \cdot J_{j} \cdot k_{q} \over s_{jq}}=\sum_{j}{(\epsilon_{q} \cdot k_{j})(\epsilon_{q} \cdot J_{j} \cdot k_{q}) \over s_{jq}}\,,~~~~\label{soft factor of sub-leading order of gravitation}
\eea
\bea
{S}_{h}^{(2)}(q) \equiv -{\tau \over 2} \sum_{j}{k_{q} \cdot J_{j} \cdot {\varepsilon}_{q} \cdot J_{j} \cdot k_{q} \over s_{jq}}={\tau \over 2} \sum_{j}{({\epsilon}_{q} \cdot J_{j} \cdot k_{q})(\epsilon_{q} \cdot J_{j} \cdot k_{q}) \over s_{jq}}\,,~~~~\label{soft factor of sub-sub-leading order of gravitation}
\eea
where ${\varepsilon}_{i}^{\mu \nu}={\epsilon}_{i}^{\mu} \epsilon_{i}^{\nu}$ are the polarization tensors carried the gravitons,
and are decomposed into two sectors of polarization vectors. The summations are for all $j \neq q$. The $s_{jq}={(k_{j}+k_{q})}^{2}=2k_{j}\cdot k_{q}=s_{qj}$ are usual Mandelstam variables.

For pure Einstein gravity, two sectors of polarization vectors are equivalent to each other. Meanwhile, for the extended gravity that Einstein gravity couples to $2$-form and dilaton fields,
each polarization tensor should be decomposed to two independent sectors of polarization vectors as $\varepsilon^{\mu\nu}_i=\epsilon_i^\mu\tilde{\epsilon}_i^\nu$. The soft theorems displayed above hold for amplitudes of Einstein gravity, rather than the extended gravity. However, since the transmutation operators which serve as the basic tool of this note requires the manifested double copy structure, as will be discussed in the next subsection, we need to label some of $\epsilon_i$ as $\tilde{\epsilon_i}$, to distinguish which Lorentz invariants will be acted by transmutation operators and which will not. This procedure will be done in section \ref{subsec-e-We}.

%%%%%%%%%%%%%%%%%%%%%%%%%%%%%%%%%%%%%%%
\subsection{Differential operators}
\label{Chapter 2, Section2}
%%%%%%%%%%%%%%%%%%%%%%%%%%%%%%%%%%%%%%%

The transmutation operators proposed in \cite{Cheung:2017ems}. There are combination of some basic differential operators act on Lorentz invariants for external momenta and polarizations. Let us list the necessary basic operators first.

The trace operator ${\cal T}_{ij}$ is defined as
\bea
{\cal T}_{ij} \equiv \frac{\partial}{\partial (\epsilon_{i} \cdot \epsilon_{j})} \equiv {\partial}_{\epsilon_{i} \epsilon_{j}}
\,,~~~~\label{trace operator}
\eea
where $\epsilon_{i} \epsilon_{j}$ means $\epsilon_{i} \cdot \epsilon_{j}$. The role of the trace operator is explained later.

The insertion operator is defined by \cite{Cheung:2017ems}.
\bea
{\cal T}_{ikj} \equiv \frac{\partial}{\partial (k_{i} \cdot \epsilon_{k})} - \frac{\partial}{\partial (k_{j} \cdot \epsilon_{k})} \equiv {\partial}_{k_{i} \epsilon_{k}} - {\partial}_{k_{j} \epsilon_{k}}
\,.~~~~\label{insertion operator}
\eea
As pointed out in \cite{Cheung:2017ems}, the ${\cal T}_{ikj}$ opertors itself is not consistent with the gauge invariance requirement. However, when it acts on objects obtained after acting one trace operators defined above, the gauge invariance is effectively protected. For an amplitude without color ordering, the effect of applying the insertion operator ${\cal T}_{(q-1)q(q+1)}$ is to insert the element $q$ between elements $q$-1 and $q$+1, which implies that the element $q$ has a relative ordering.

Using the products of basic operators, more complicated operators can be constructed. In this paper, we will employ the combinatorial operators which connect GR and YM amplitudes together, and connect YM and BAS amplitudes in the same manner. Of course, the product of basic operators can also link Einstein-Yang-Mills (EYM) and GR amplitudes together, and connect Yang-Mills-scalar (YMS) and YM amplitudes in the same manner \cite{Cheung:2017ems,Zhou:2018wvn,Bollmann:2018edb,Tao:2024vcz}.

Such combinatorial operator ${\cal T}[\sigma]$ is defined as \cite{Cheung:2017ems}
\bea
{\cal T}[\sigma] \equiv {\cal T}_{1n} \prod \limits_{i=2}^{n-1}{\cal T}_{(i-1)in}
\,,~~~~\label{composed operator 1}
\eea
where we uses $\sigma$ to denote the ordered set $\sigma=\{1,\cdots,n\}$ throughout this paper.
The effect of acting ${\cal T}[\sigma]$ on the tree level amplitude is to reduce the spin of all external particles by $1$, and generate a new ordering.
The creation of the ordering $\sigma=\{1,\cdots,n\}$ can be understood as follows. The trace operators ${\cal T}_{1n}$ generates two ends of the ordering, then the insertion operator ${\cal T}_{12n}$ creates $(1,2,n)$ by inserting $2$ between $1$ and $n$. Repeating the above interpretation of the effect of insertion operator, one can achieve the full ordering ultimately.

The corresponding transmutation relation is given by \cite{Cheung:2017ems}
\bea
{\cal A}_{\rm YM}(\sigma)={\cal T}[\sigma]{\cal A}_{\rm G}(\sigma_{no})
\,,~~~~\label{relation between Aym and Ag}
\eea
\bea
{\cal A}_{\rm BAS}(\sigma|\sigma^{'})={\cal T}[\sigma]{\cal A}_{\rm YM}(\sigma^{'})
\,,~~~~\label{relation between Abas and Aym}
\eea
where $\sigma^{'}$ is an arbitrary ordered set for elements in $(1,\cdots,n)$. Note that \eref{relation between Aym and Ag} only holds for extended gravity. ${\cal A}_{\rm YM}(\sigma)$ represents the $n$-point tree YM amplitudes with one ordering $\sigma$. ${\cal A}_{\rm BAS}(\sigma|\sigma^{'})$ represents the $n$-point double ordered tree BAS amplitudes which carry two independent orderings $\sigma$ and $\sigma^{'}$. The first example, the result of ${\cal T}[1,2,3,4,5]={\cal T}_{15}{\cal T}_{125}{\cal T}_{235}{\cal T}_{345}$ acting on the 5-point gravity amplitude is a 5-point YM amplitude with a ordering $\sigma=\{1,2,3,4,5\}$, namely, ${\cal A}_{\rm YM}(1,2,3,4,5)={\cal T}[1,2,3,4,5]{\cal A}_{\rm G}(5-point)$. The second example, the result of ${\cal T}[1,3,2,5,4]={\cal T}_{14}{\cal T}_{134}{\cal T}_{324}{\cal T}_{254}$ acting on the 5-point YM amplitude ${\cal A}_{\rm YM}(1,2,3,4,5)$ is a 5-point BAS amplitude which carry two independent orderings $\sigma=\{1,3,2,5,4\}$ and $\sigma^{'}=\{1,2,3,4,5\}$, namely, ${\cal A}_{\rm BAS}(1,3,2,5,4|1,2,3,4,5)={\cal T}[1,3,2,5,4]{\cal A}_{\rm YM}(1,2,3,4,5)$.

In fact, there are more than one allowed formulas of combinatorial operators those satisfy $\eref{relation between Aym and Ag}$ and $\eref{relation between Abas and Aym}$. First, one can use momentum conservation to replace $k_{n}$ with $-\sum_{i \neq n}k_{i}$, then the combinatorial operator ${\cal T}[\sigma]$ becomes
\bea
{\cal T}_{1}[\sigma] \equiv {\cal T}_{1n} \prod \limits_{i=2}^{n-1}\partial_{k_{i-1} \epsilon_{i}}
\,.~~~~\label{composed operator 1 1}
\eea
Replacing ${\cal T}[\sigma]$ with ${\cal T}_{1}[\sigma]$, \eref{relation between Aym and Ag} and \eref{relation between Abas and Aym} are also true. Furthermore, based on the interpretation of the effect of insertion operator, one can also chose
\bea
{\cal T}_{0}[\sigma] \equiv {\cal T}_{1n} \prod \limits_{i \neq 1,q,q+1,n} \partial_{k_{i-1} \epsilon_{i}} \partial_{k_{q-1} \epsilon_{q+1}} {\cal T}_{(q-1)q(q+1)}={\cal T}[\sigma/q] {\cal T}_{(q-1)q(q+1)}
\,,~~~~\label{composed operator 1 0}
\eea
with
\bea
{\cal T}[\sigma/q]={\cal T}_{1n} \prod \limits_{i \neq 1,q,q+1,n} \partial_{k_{i-1} \epsilon_{i}}\partial_{k_{q-1} \epsilon_{q+1}}\,.~~\label{defin-T-sigma/q}
\eea
This operator has the same effect as ${\cal T}_1[\sigma]$ when acting on physical amplitudes. The alternative operator ${\cal T}_0[\sigma]$ creates the ordering $\sigma/q=\{1,\cdots,q-1,q+1,\cdots,n\}$ first, then insert the leg $q$ between $(q-1)$ and $(q+1)$. For example, we set $q=3$, the result of ${\cal T}_{0}[1,2,3,4,5]={\cal T}_{15}\partial_{k_{1} \epsilon_{2}}\partial_{k_{2} \epsilon_{4}}{\cal T}_{234}={\cal T}[1,2,4,5]{\cal T}_{234}$ acting on the 5-point gravity amplitude is a 5-point YM amplitude with a ordering $\sigma=\{1,2,3,4,5\}$, namely, ${\cal A}_{\rm YM}(1,2,3,4,5)={\cal T}_{0}[1,2,3,4,5]{\cal A}_{\rm G}(5-point)$. We can clearly see that ${\cal T}[1,2,3,4,5] \neq {\cal T}_{0}[1,2,3,4,5]$, but they all result in the same YM amplitude ${\cal A}_{\rm YM}(1,2,3,4,5)$ when acted to the 5-point gravity amplitude. 

So far, we have seen that differential operators or transmutation operators link BAS, YM, and gravity amplitude in tree levle. Next, let's take a look at the relationship between the expansion coefficient of amplitude and differential operators.

%%%%%%%%%%%%%%%%%%%%%%%%%%%%%%%%%%%%%%%
\subsection{Expansions of amplitudes}
\label{Chapter 2, Section3}
%%%%%%%%%%%%%%%%%%%%%%%%%%%%%%%%%%%%%%%

The action of operator ${\cal T}[\sigma]$ can be further understood via the well known Cachazo-He-Yuan (CHY) formula \cite{Cachazo:2014xea,Cachazo:2013hca,Cachazo:2013iea}. In CHY formula,
the corresponding integrand for GR, YM, and BAS theories are given as
\bea
{\cal I}_{\rm GR}&=&\mathrm{Pf'}\Psi(\epsilon,k)\,\mathrm{Pf'}\tilde{\Psi}(\tilde{\epsilon},k)\,,\nn
{\cal I}_{\rm YM}(\sigma)&=&\mathrm{Pf'}\Psi(\epsilon,k)\,\mathrm{PT}(\sigma)\,,\nn
{\cal I}_{\rm BAS}(\sigma|\sigma')&=&\mathrm{PT}(\sigma)\,\mathrm{PT}(\sigma')\,,
\eea
where $\mathrm{Pf'}\Psi(\epsilon,k)$ stands for a polynomial depends on external polarizations $\epsilon_i$ and momenta $k_i$, and is called the reduced Pfaffian of the matrix $\Psi(\epsilon,k)$. In the above, the GR integrand is for the extended one, with independent polarizations $\epsilon_i$ and $\tilde{\epsilon}_i$ included in two reduced Pfaffians, respectively. Of course, such CHY integrand also holds for Einstein gravity, with the equivalent $\epsilon_i$ and $\tilde{\epsilon}_i$. The block $\mathrm{PT}(\sigma)$ is the so called Parke-Taylor factor with ordering $\sigma$, which is independent of any external kinematic variable.
As discussed in \cite{Zhou:2018wvn,Bollmann:2018edb}, the effect of acting the combinatorial operator ${\cal T}[\sigma]$ on $\mathrm{Pf'}\Psi(\epsilon,k)$ is
\bea
{\cal T}[\sigma] \mathrm{Pf'}\Psi(\epsilon,k)=\mathrm{PT}(\sigma)
\,,~~~~\label{PT0}
\eea
this relation immediately gives rise to \eref{relation between Aym and Ag} and \eref{relation between Abas and Aym}.

The reduced Pfaffian can be expanded to Parke-Taylor factors as
\bea \mathrm{Pf'}\Psi(\epsilon,k)=\sum_{\sigma} C(\epsilon,k;\sigma)\mathrm{PT}(\sigma)
\,,~~~~\label{C and PT}
\eea
where the coefficients $C(\epsilon,k;\sigma)$ depend on polarizations, momenta and the ordering $\sigma$, and have mass dimension $n-2$.
This expansion in \eref{C and PT} indicates the following expansions of amplitudes
\bea
{\cal A}_{\rm G}(\sigma_{no})&=&\sum_{\sigma}\,\tilde{C}(\tilde{\epsilon},k;\sigma)\,{\cal A}_{\rm YM}(\sigma)=\sum_{\sigma}\,C(\epsilon,k;\sigma)\,\tilde{{\cal A}}_{\rm YM}(\sigma)\,,\nn
{\cal A}_{\rm YM}(\sigma)&=&\sum_{\sigma'}\,{C}({\epsilon},k;\sigma')\,{\cal A}_{\rm BAS}(\sigma|\sigma')
\,.~~\label{expand G and YM}
\eea
Combining \eref{PT0} and \eref{C and PT} together, we can also understand the effect of ${\cal T}[\sigma]$ as
\bea
{\cal T}[\sigma]\, C(\epsilon,k;\sigma')=\delta_{\sigma \sigma'}
\,.~~~~\label{T act PT}
\eea

The explicit forms of $\mathrm{Pf'}\Psi(\epsilon,k)$, $\mathrm{PT}(\sigma)$ and $C(\epsilon,k;\sigma)$ are irrelevant for us. On the other hand,
the expansion \eref{expand G and YM} and relation \eref{T act PT} will be used in next section.

%%%%%%%%%%%%%%%%%%%%%%%%%%%%%%%%%%%%
\section{From the soft theorems of gravity to those of Yang-Mills}
\label{Chapter 3}
%%%%%%%%%%%%%%%%%%%%%%%%%%%%%%%%%%%%

The purpose of this section is to investigate the soft behavior of YM theory from those of GR, by using transmutation operators introduced in section \ref{Chapter 2}. Of course, each $n$-point amplitude ${\cal A}_n$ can be expanded by the soft parameter $\tau$ as
\bea
{\cal A}_n(\tau)=\sum_{a=0}^\infty\,\tau^{a+\ell}\,{\cal A}^{(a)}_n\,,~~\label{expand by tau}
\eea
if we take $k_q\to\tau k_q$ and $\tau\to 0$, where $\ell$ is an integer, which takes the value $\ell=-1$ for the GR case. We are interested in the question that whether the component ${\cal A}^{(a)}_n$ at $a^{\rm th}$ order
can factorize as
\bea
\tau^{a+\ell}{\cal A}^{(a)}_n=S^{(a)}(q)\,{\cal A}_{n-1}\,,~~\label{factorize}
\eea
where $S^{(a)}(q)$ is an universal soft factor valid for arbitrary number of external legs, and ${\cal A}_{n-1}$ stands for the $(n-1)$-point amplitude, generated from ${\cal A}_n$ by removing the soft leg $q$. Since the transmutation relation \eref{relation between Aym and Ag} links GR and YM amplitudes together,
it is natural to study the above factorization in \eref{factorize} for YM amplitudes via such connection.
We will derive soft behaviors of YM amplitudes at each order by using ${\cal T}_{0}[\sigma]$ given in \eref{composed operator 1 0}, then use ${\cal T}_{1}[\sigma]$ in \eref{composed operator 1 1} to verify the result.

%%%%%%%%%%%%%%%%%%%%%%%
\subsection{General discussion on effects of $\tilde{\cal T}_0[\sigma]$ and $\tilde{\cal T}_1[\sigma]$}
\label{subsec-e-We}
%%%%%%%%%%%%%%%%%%%%%%%%%%%%

When considering both the soft behavior of the amplitude and combinatorial operators acting on the amplitude, $\partial_{k_{q}\epsilon_{j}}$ in the insertion operator is simply replaced by ${1 \over \tau} \partial_{k_{q}\epsilon_{j}}$. Thus, ${\cal T}_{1}[\sigma]$ is proportional to $\tau^{-1}$ (${\cal T}_{1}[\sigma] \propto \tau^{-1}$), ${\cal T}_{0}[\sigma]$ is proportional to $\tau^{0}$ (${\cal T}_{0}[\sigma] \propto \tau^{0}$, independent of $\tau$). One can also chose combinatorial operators those contain different parts at different order of $\tau$. Obviously, operators at different order play different roles when connecting soft behaviors of GR and YM amplitudes (or YM and BAS amplitudes).

We will use transmutation operators $\tilde{\cal T}_0[\sigma]$ and $\tilde{\cal T}_1[\sigma]$, defined in \eref{composed operator 1 0} and \eref{composed operator 1 1}, to connect soft behaviors of GR and YM amplitudes. Before doing the detailed calculations, let us give the general discussion on the effects of these two operators, which are related to soft behaviors. By definition, the operator $\tilde{\cal T}_0[\sigma]$ is independent of the soft momentum $k_q$, thus does not carry the scale parameter $\tau$. Then, by expanding GR and YM amplitudes as in \eref{expand by tau}, we see the operator $\tilde{\cal T}_0[\sigma]$ transmutes the GR soft behavior at $a^{\rm th}$ order as
\bea
{\cal A}^{(a)}_{\rm YM}(\sigma)=\tilde{{\cal T}}_0[\sigma]{\cal A}^{(a)}_{\rm G}(\sigma_{no})\,,~~\label{effect-T0}
\eea
based on the relation \eref{relation between Aym and Ag}. Meanwhile, the operator $\tilde{\cal T}_1[\sigma]$ acts on $\tilde{\epsilon}_{q+1}\cdot k_q$, which is accompanied with $\tau$. It means this operator transmutes GR terms as
\bea
{\cal A}^{(a)}_{\rm YM}(\sigma)=\tilde{{\cal T}}_1[\sigma]{\cal A}^{(a+1)}_{\rm G}(\sigma_{no})\,.~~\label{effect-T1}
\eea

As pointed out in section \ref{Chapter 2, Section2}, the transmutation operator $\tilde{{\cal T}}[\sigma]$  defined with polarizations $\tilde{\epsilon}_i$, only acts on $\mathrm{Pf'}\tilde{\Psi}(\tilde{\epsilon},k)$ or $\tilde{C}(\tilde{\epsilon},k;\sigma)$, which involves $\tilde{\epsilon}_i$.
However, the soft factors displayed in \eref{soft factor of leading order of gravitation}, \eref{soft factor of sub-leading order of gravitation}, \eref{soft factor of sub-sub-leading order of gravitation} do not distinguish $\epsilon_i$ and $\tilde{\epsilon}_i$. Although the GR theory under consideration is the standard Einstein gravity, in order to perform transmutation operators, we need to recognize the $\tilde{\epsilon}_i$ part under the action of $\tilde{\cal T}[\sigma]$ (or equivalently the $\epsilon_i$ part under the action of ${\cal T}[\sigma]$).

Clearly, the leading soft factor in \eref{soft factor of leading order of gravitation} can be rewritten as
\bea
{S}_{h}^{(0)}(q)={1 \over \tau} \sum_{j} {({\epsilon}_{q} \cdot k_{j})(\tilde{\epsilon}_{q} \cdot k_{j}) \over s_{jq}}
\,,~~~~\label{soft factor of leading order of extended gravity}
\eea
due to the direct decomposition $\varepsilon_q^{\mu\nu}=\epsilon_q^\mu\tilde{\epsilon}_q^\nu$. The expansion of GR amplitudes in \eref{expand G and YM} indicates that the leading soft behavior reads
\bea
{\cal A}_{\rm G}^{(0)}(\sigma_{no})=\sum_{\sigma}\tilde{C}^{(0)}(\tilde{\epsilon},k;\sigma){\cal A}_{\rm YM}^{(0)}(\sigma)
\,,~~\label{Ag0 expand Aym}
\eea
where $\tilde{C}^{(0)}(\tilde{\epsilon},k;\sigma)$ encodes the leading contribution of $\tilde{C}(\tilde{\epsilon},k;\sigma)$, while ${\cal A}_{\rm YM}^{(0)}(\sigma)$ denotes the leading contribution of ${\cal A}_{\rm YM}(\sigma)$.
Thus, the leading soft factor in \eref{soft factor of leading order of extended gravity} can be understood as, the factor $\tilde{\epsilon}_{q} \cdot k_{j}$ arises from coefficients $\tilde{C}^{(0)}(\tilde{\epsilon},k;\sigma)$, while the remaining ${\epsilon}_{q} \cdot k_{j}/s_{jq}$ is contributed by YM amplitudes ${\cal A}_{\rm YM}^{(0)}(\sigma)$.

The sub-leading order is more subtle. From the expansion in \eref{expand G and YM}, one get the following sub-leading soft behavior
\bea
{\cal A}_{\rm G}^{(1)}(\sigma_{no})=\sum_{\sigma}\tilde{C}^{(0)}(\tilde{\epsilon},k;\sigma){\cal A}_{\rm YM}^{(1)}(\sigma)+\sum_{\sigma}\tilde{C}^{(1)}(\tilde{\epsilon},k;\sigma){\cal A}_{\rm YM}^{(0)}(\sigma)
\,.~~\label{Ag1 expand Aym}
\eea
From the above expansion, one can extract sub-leading or leading soft behaviors of YM amplitudes, namely ${\cal A}_{\rm YM}^{(1)}(\sigma)$
or ${\cal A}_{\rm YM}^{(0)}(\sigma)$, by acting different transmutation operators.
The first case, the operator $\tilde{\cal T}_0[\sigma]$ transmutes ${\cal A}_{\rm G}^{(1)}(\sigma_{no})$ to ${\cal A}_{\rm YM}^{(1)}(\sigma)$,
corresponds to $a=1$ in \eref{effect-T0}. The second case, the operator $\tilde{\cal T}_1[\sigma]$ transmutes ${\cal A}_{\rm G}^{(1)}(\sigma_{no})$ to ${\cal A}_{\rm YM}^{(0)}(\sigma)$, corresponds to $a=0$ in \eref{effect-T1}. These extractions are ensured by the relations
\bea
& &\tilde{\cal T}_0[\sigma']\tilde{C}^{(0)}(\tilde{\epsilon},k;\sigma)=\delta_{\sigma'\sigma}\,,\nn
& &\tilde{\cal T}_0[\sigma']\tilde{C}^{(a)}(\tilde{\epsilon},k;\sigma)=0\,,~~~~{\rm for}~a\neq 0\,,~~\label{require1}
\eea
and
\bea
& &\tilde{\cal T}_1[\sigma']\tilde{C}^{(a)}(\tilde{\epsilon},k;\sigma)=0\,,~~~~{\rm for}~a\neq1\,,\nn
& &\tilde{\cal T}_1[\sigma']\tilde{C}^{(1)}(\tilde{\epsilon},k;\sigma)=\delta_{\sigma'\sigma}\,.~~\label{require2}
\eea
The above properties can be verified straightforwardly. The leading coefficient $\tilde{C}^{(0)}(\tilde{\epsilon},k;\sigma)$ should be independent of $k_q$ since $k_q$ is accompanied with $\tau$, and the sub-leading coefficient $\tilde{C}^{(1)}(\tilde{\epsilon},k;\sigma)$ should be linear in $\tau$. Such feature can be directly extended to arbitrary $\tilde{C}^{(a)}(\tilde{\epsilon},k;\sigma)$ which are multi-linear in $k_q$. Therefore, the operator $\tilde{\cal T}_0[\sigma']$ which do not involve $k_q$, annihilates each $\tilde{C}^{(a)}(\tilde{\epsilon},k;\sigma)$ with $a\neq0$. Similarly, the operator $\tilde{\cal T}_1[\sigma']$ which acts on
$k_q\cdot\tilde{\epsilon}_{q+1}$, annihilates each $\tilde{C}^{(a)}(\tilde{\epsilon},k;\sigma)$ with $a\neq1$. Then the requirements \eref{require1} and \eref{require2} are ensured by \eref{T act PT}.
Based on the above expectation,
it is natural to split the sub-leading soft factor in \eref{soft factor of sub-leading order of gravitation} as
\bea
{S}_{h}^{(1)}(q)=\sum_{j}{(\tilde{\epsilon}_{q} \cdot k_{j})(\epsilon_{q} \cdot L_{j} \cdot k_{q}) \over s_{jq}}+\sum_{j}{(\epsilon_{q} \cdot k_{j})(\tilde{\epsilon}_{q} \cdot \tilde{L}_{j} \cdot k_{q}) \over s_{jq}}
\,.~~~~\label{soft factor of sub-leading order of extended gravity}
\eea
Here the operator $L_j$ acts on external polarizations and momenta carried by YM amplitudes in \eref{expand G and YM} as
\bea
L_j^{\mu\nu}\,k_j^\rho= k_j^{\mu}\,
{\partial k_j^\rho\over\partial k_{j,\nu}}-k_j^{\nu}\,{\partial k_j^\rho\over\partial k_{j,\mu}}\,,~~~~
L_j^{\mu\nu}\,\epsilon_j^\rho=\big(\eta^{\nu\rho}\,\delta^\mu_\sigma-\eta^{\mu\rho}\,\delta^\nu_\sigma\big)\,
\epsilon^\sigma_j\,,~~\label{L-left}
\eea
but never acts on coefficients $\tilde{C}(\tilde{\epsilon},k;\sigma)$. On the other hand, the operator $\tilde{L}_{j}$
only acts on polarizations and momenta in $\tilde{C}(\tilde{\epsilon},k;\sigma)$, with the effect analogous to \eref{L-left}.
The splitting in \eref{soft factor of sub-leading order of extended gravity} breaks the manifest symmetry among $\epsilon_i$ and $\tilde{\epsilon}_i$, however, since $\epsilon_i$ and $\tilde{\epsilon}_i$ are equivalent to each other in Einstein gravity, this phenomenon does not violate any physical condition. Using the observation that $\tilde{{C}}(\tilde{\epsilon},k;\sigma)$ are polynomials depend on $\tilde{\epsilon}_i$, without any pole, while ${\cal A}_{\rm YM}(\sigma)$ depend on $\epsilon_i$, one can understand two terms in \eref{soft factor of sub-leading order of extended gravity} as follows. The factor $\tilde{\epsilon}_{q} \cdot k_{j}$ arises from $\tilde{C}^{(0)}(\tilde{\epsilon},k;\sigma)$, as already figured out in the leading soft factor \eref{soft factor of leading order of extended gravity}. It means this term corresponds to the first term  in \eref{Ag1 expand Aym}, thus the factor $\epsilon_{q} \cdot L_{j} \cdot k_{q}/ s_{jq}$ is contributed by ${\cal A}_{\rm YM}^{(1)}(\sigma)$. Similar argument shows that, in the second term, $\tilde{\epsilon}_{q} \cdot \tilde{L}_{j} \cdot k_{q}$ comes from $\tilde{C}^{(1)}(\tilde{\epsilon},k;\sigma)$, while $\epsilon_{q} \cdot k_{j}/ s_{jq}$ is provided by ${\cal A}_{\rm YM}^{(0)}(\sigma)$.

The most complicated case is the sub-sub-leading order. The expansion in \eref{expand G and YM} tells us
\bea
{\cal A}_{\rm G}^{(2)}(\sigma_{no})&=&\sum_{\sigma}\tilde{C}^{(0)}(\tilde{\epsilon},k;\sigma){\cal A}_{\rm YM}^{(2)}(\sigma)+\sum_{\sigma}\tilde{C}^{(1)}(\tilde{\epsilon},k;\sigma){\cal A}_{\rm YM}^{(1)}(\sigma)\nn
& &+\sum_{\sigma}\tilde{C}^{(2)}(\tilde{\epsilon},k;\sigma){\cal A}_{\rm YM}^{(0)}(\sigma)
\,,~~\label{Ag2 expand Aym}
\eea
thus one can use $\tilde{\cal T}_0[\sigma]$ and $\tilde{\cal T}_1[\sigma]$  to extract ${\cal A}_{\rm YM}^{(2)}(\sigma)$ and ${\cal A}_{\rm YM}^{(1)}(\sigma)$ respectively, based on relations in \eref{require1} and \eref{require2}.
The difficulty is how to represent the complete ${\cal A}_{\rm G}^{(2)}(\sigma_{no})$ in the form that $\epsilon_i$ and $\tilde{\epsilon}$ are correctly distinguished. Fortunately, for our purpose, this problem can be bypassed.
For $a=1$ in \eref{effect-T1}, the effective part is $\tilde{C}^{(1)}(\tilde{\epsilon},k;\sigma){\cal A}_{\rm YM}^{(1)}(\sigma)$ part, thus the soft operator which is needed for this case, is
the subpart of the full one \eref{soft factor of sub-sub-leading order of gravitation}, i.e.,
\bea
{\cal S}_{h}^{(2)}(q)=\sum_{j}{\tau\,({\epsilon}_{q} \cdot L_{j} \cdot k_{q})(\tilde{\epsilon}_{q} \cdot \tilde{L}_{j} \cdot k_{q}) \over s_{jq}}\,,~~~~\label{soft factor of sub-sub-leading order of extended GR-partial}
\eea
with $L_{j}$ and $\tilde{L}_{j}$ defined previously. Here we used the notation ${\cal S}$ to emphasize that this operator is not the complete one. Due to the previous discussion, $\tilde{\epsilon}_{q} \cdot \tilde{L}_{j} \cdot k_{q}$ arises from $\tilde{C}^{(1)}(\tilde{\epsilon},k;\sigma)$,
and $\epsilon_{q} \cdot L_{j} \cdot k_{q}/ s_{jq}$ corresponds to ${\cal A}_{\rm YM}^{(1)}(\sigma)$. For $a=2$ in \eref{effect-T0}, we can explain the reason why ${\cal A}_{\rm YM}^{(2)}(\sigma)$ can not factorize into an universal soft factor and a $(n-1)$-point amplitude, without exploring the explicit form of the corresponding soft factor.

\subsection{From GR to YM: operator ${\cal T}_{0}[\sigma]$}
\label{Chapter 3, Section 1}
%%%%%%%%%%%%%%%%%%%%%%%%%%%%%%%%%%%%

Now we study the soft behaviors of YM amplitudes, by acting ${\cal T}_{0}[\sigma]$ on GR terms at different orders.
%The following helpful properties will be used.
%First, both GR and YM amplitudes are linear in each polarization, namely, $\partial{\cal A}_{\rm G}/\partial\epsilon_i^\mu$ and $\partial{\cal %A}_{\rm YM}/\partial\epsilon_i^\mu$ are independent of $\epsilon_i^\mu$, and the situation for $\tilde{\epsilon}_i^\mu$ is analogous.
%Secondly,
Due to the definition of operator $L_j$ in \eref{L-left}, we have the following useful property when acting $L_j$ on physical amplitudes,
\bea
{\epsilon}_{q} \cdot L_{j} \cdot k_{q}
&=&-(k_{j} \cdot f_{q} \cdot \partial_{k_{j}}+\epsilon_{j} \cdot f_{q} \cdot \partial_{\epsilon_{j}})
\,,~~~~\label{relation of J and f}
\eea
where the linearity on each polarization is used.
The situation for $\sum_{j}\tilde{\epsilon}_{q} \cdot \tilde{L}_{j} \cdot k_{q}$ is analogous, with $f$ replaced by $\tilde{f}$. Using \eref{relation of J and f}, we can easily verify the following equation
\bea
\tilde{\cal T}_{0}[\sigma] \sum_{j}\left(\tilde{\epsilon}_{q} \cdot \tilde{L}_{j} \cdot k_{q}\right)\sum_{\sigma/q}\tilde{C}(\tilde{\epsilon},k;\sigma/q)=0
\,,~~\label{T0 L C}
\eea
where $\tilde{C}(\tilde{\epsilon},k;\sigma/q)$ serve as coefficient of expansion \eref{expand G and YM} for ${\cal A}_{\rm G}(\sigma_{no}/q)$ obtained from ${\cal A}_{\rm G}(\sigma_{no})$ by taking $k_q=0$ and removing the leg $q$.

\textbf{1}. Calculate $\tilde{\cal T}_{0}[\sigma] {\cal A}_{\rm G}^{(0)}(\sigma_{no})$ :

According to \eref{effect-T0}, when $a=0$, we have
\bea
{\cal A}_{\rm YM}^{(0)}(\sigma)&=&\tilde{\cal T}_{0}[\sigma] {\cal A}_{\rm G}^{(0)}(\sigma_{no})\nn
&=&\left(\tilde{\cal T}_{(q-1)q(q+1)}\,{S}_{h}^{(0)}(q)\right) \tilde{\cal T}[\sigma/q]{\cal A}_{\rm G}(\sigma_{no}/q)\nn
&=&{1 \over \tau}\left(\tilde{\cal T}_{(q-1)q(q+1)} \sum_{j} {(\epsilon_{q} \cdot k_{j})(\tilde{\epsilon}_{q} \cdot k_{j}) \over s_{jq}}\right){\cal A}_{\rm YM}(\sigma/q)\nn
&=&{1 \over \tau} \left({\epsilon_{q} \cdot k_{q-1} \over s_{q(q-1)}}-{\epsilon_{q} \cdot k_{q+1} \over s_{q(q+1)}}\right){\cal A}_{\rm YM}(\sigma/q)
\,,~~~~\label{composed operator 1 0 acting on Ag0}
\eea
where the first equality uses \eref{composed operator 1 0}. In the second equality, we used the leading soft factor of GR amplitudes in \eref{soft factor of leading order of extended gravity}. The last step uses ${\cal A}_{\rm YM}(\sigma/q)=\tilde{\cal T}[\sigma/q] {\cal A}_{\rm G}(\sigma_{no}/q)$, with $\tilde{\cal T}[\sigma/q]$ defined in \eref{defin-T-sigma/q}. The result in \eref{composed operator 1 0 acting on Ag0} is coincide with
\cite{Arkani-Hamed:2008owk,Casali:2014xpa,Zhou:2022orv}. This result indicates that the YM amplitude satisfies the soft theorem at the leading order (proportional to $\tau^{-1}$).

\textbf{2}. Calculate $\tilde{\cal T}_{0}[\sigma] {\cal A}_{\rm G}^{(1)}(\sigma_{no})$ :

According to \eref{effect-T0}, when $a=1$, we have
\bea
{\cal A}_{\rm YM}^{(1)}(\sigma)&=&\tilde{\cal T}_{0}[\sigma] {\cal A}_{\rm G}^{(1)}(\sigma_{no}/q)\nn
&=&\sum_{j}\Big[\tilde{\cal T}_{(q-1)q(q+1)}(\tilde{\epsilon}_{q} \cdot k_{j}) \Big]{\epsilon_{q} \cdot L_{j} \cdot k_{q} \over s_{jq}}\tilde{\cal T}[\sigma/q]{\cal A}_{\rm G}(\sigma_{no}/q)\nn
&=&\left({\epsilon_{q} \cdot L_{q-1} \cdot k_{q} \over s_{q(q-1)}}-{\epsilon_{q} \cdot L_{q+1} \cdot k_{q} \over s_{q(q+1)}}\right){\cal A}_{\rm YM}(\sigma/q)\nn
&=&\left({\epsilon_{q} \cdot J_{q-1} \cdot k_{q} \over s_{q(q-1)}}-{\epsilon_{q} \cdot J_{q+1} \cdot k_{q} \over s_{q(q+1)}}\right){\cal A}_{\rm YM}(\sigma/q)
\,,~~~~\label{composed operator 1 0 acting on Ag1}
\eea
where we have used the first term in the split soft operator \eref{soft factor of sub-leading order of extended gravity}, which corresponds to the first term in \eref{Ag1 expand Aym}.
The last step uses the observation that for the pure YM case operators $L_{q-1}$ and $L_{q+1}$ are equivalent to angular momentum operators $J_{q-1}$ and $J_{q+1}$. The result in \eref{composed operator 1 0 acting on Ag1} is coincide with that in
\cite{Arkani-Hamed:2008owk,Casali:2014xpa,Zhou:2022orv}. This result indicates that the YM amplitude satisfies the soft theorem at the sub-leading order (proportional to $\tau^{0}$).

\textbf{3}. $\tilde{\cal T}_{0}[\sigma] {\cal A}_{\rm G}^{(2)}(\sigma_{no})$ :

The expansion \eref{Ag2 expand Aym} indicates that acting $\tilde{\cal T}_{0}[\sigma]$ on ${\cal A}_{\rm G}^{(2)}(\sigma_{no})$
gives arise to the sub-sub-leading soft behavior ${\cal A}_{\rm YM}^{(2)}(\sigma)$. However, it is not easy to determine the full
${\cal A}_{\rm G}^{(2)}(\sigma_{no})$ with $\epsilon_i$ and $\tilde{\epsilon}_i$ are distinguished explicitly. But we can find that
at the sub-sub-leading order, the YM soft behavior ${\cal A}_{\rm YM}^{(2)}(\sigma)$ can never factorize into an universal soft behavior and the $(n-1)$-point sub-amplitude, due to the speciality of the $4$-point case.

Let us act the soft factor \eref{soft factor of sub-sub-leading order of gravitation} on the $(n-1)$-point GR amplitude, without distinguishing $\epsilon_i$ and $\tilde{\epsilon}_i$ at this first step. When acting on the Lorentz invariant $k_{q-1}\cdot k_{q+1}$, we get
\bea
S^{(2)}_h(q)(k_{q-1}\cdot k_{q+1})=(k_{q-1}\cdot f_q\cdot k_{q+1})\Big({\epsilon_q\cdot J_{q+1}\cdot k_q\over s_{(q+1)q}}-{\epsilon_q\cdot J_{q-1}\cdot k_q\over s_{(q-1)q}}\Big)\,,~~\label{s2-on-kk}
\eea
where the property \eref{relation of J and f} has been used (with $L_j$ replaced by $J_j$). To identify the contribution from $\tilde{C}^{(0)}(\tilde{\epsilon},k;\sigma)$, we should turn $\epsilon_q$ in $f_q$ to $\tilde{\epsilon}_q$, and obtain
\bea
\tilde{{\cal T}}_{(q-1)q(q+1)}\Big(S^{(2)}_h(q)(k_{q-1}\cdot k_{q+1})\Big)&=&{(k_q\cdot k_{q+1})(\epsilon_q\cdot J_{q-1}\cdot k_q)\over s_{q(q-1)}}
-{(k_q\cdot k_{q-1})(\epsilon_q\cdot J_{q+1}\cdot k_q)\over s_{q(q+1)}}\nn
& &+{\epsilon_q\cdot J_{q-1}\cdot k_q\over2}-{\epsilon_q\cdot J_{q+1}\cdot k_q\over2}\,,~~\label{term without pole}
\eea
where two terms without any pole occur in the second line. Obviously, the cancelation of  poles can be extended to all $v\cdot k_{q-1}$ or $v\cdot k_{q+1}$,
where $v$ stands for an arbitrary Lorentz vector. Such phenomenon holds for $n$-point amplitudes with $n\geq5$, but is violated for the $n=4$ case. The reason is, the sub-sub-leading soft term at $\tau$ order should be linear in $k_q$. On the other hand, each $4$-point YM amplitude has mass dimension $0$. Combining the above two requirements together, one find that the sub-sub-leading contribution without any pole can never exist for the $4$-point case. Based on the difference between $n=4$ and $n\geq5$ cases, we conclude that the universal soft behavior of YM amplitudes, which is valid for arbitrary number of external gluons, does not exist at the sub-sub-leading order.

The soft behaviors represented in \eref{composed operator 1 0 acting on Ag0} and \eref{composed operator 1 0 acting on Ag1} are derived for the special ordering $\sigma=\{1,2,\cdots,n\}$. It is straightforward to extend them to the general ordering $\sigma'$ as
\bea
{\cal A}_{\rm YM}^{(0)}(\sigma')&=&S^{(0)}_g(q){\cal A}_{\rm YM}(\sigma'/q)\,,\nn
{\cal A}_{\rm YM}^{(1)}(\sigma')&=&S^{(1)}_g(q){\cal A}_{\rm YM}(\sigma'/q)\,,~~~~\label{summary of soft factor of Aym}
\eea
with soft factors
\bea
{S}_{g}^{(0)}(q)={1 \over \tau} \sum_{j} {\epsilon_{q} \cdot k_{j} \over s_{jq}} \Delta_{jq}
\,,~~~~\label{leading-order soft factor of Aym}
\eea
\bea
{S}_{g}^{(1)}(q)=\sum_{j} {\epsilon_{q} \cdot J_{j} \cdot k_{q} \over s_{jq}} \Delta_{jq}
\,.~~~~\label{sub-leading-order soft factor of Aym}
\eea
Here the symbol $\Delta_{jq}$ describe the adjacency of two legs $j$ and $q$ in the ordering $\sigma'$,
\bea
\Delta_{jq}=\left\{ \begin{array}{l}
	\begin{matrix}
		+1\,,~~~~\sigma'=\{\cdots ,j,q,\cdots \}\\
	\end{matrix}\\
	\begin{matrix}
		-1\,,~~~~\sigma'=\{ \cdots ,q,j,\cdots \}\\
	\end{matrix}\\
	\begin{matrix}
		0\,,~~~~\sigma'=\{ \cdots ,j,\cdots ,q,\cdots \}~~{\rm or}~~\sigma'=\{ \cdots ,q,\cdots ,j,\cdots \}\\
	\end{matrix}\\
\end{array} \right.
\,.~~~~\label{Delta definition}
\eea
It is direct to observe that $\Delta_{jq}=-\Delta_{qj}$. The YM amplitude carries a color order, so the specific form of the soft factor is also related to this order. When $\sigma^{'}=\sigma=\{1,\cdots,q-1,q,q+1,\cdots,n\}$, \eref{summary of soft factor of Aym} returns to \eref{composed operator 1 0 acting on Ag0} and \eref{composed operator 1 0 acting on Ag1}.

%%%%%%%%%%%%%%%%%%%%%%%%%%%%%%%%%%%%
\subsection{From GR to YM: operator $\tilde{\cal T}_1[\sigma]$}
\label{Chapter 3, Section 2}
%%%%%%%%%%%%%%%%%%%%%%%%%%%%%%%%%%%%

Our goal in this section is to use combinatorial operator $\tilde{\cal T}_{1}[\sigma] \propto \tau^{-1}$ to verify the results of section \ref{Chapter 3, Section 1}. We will frequently use the observation
\bea
\partial_{k_{q-1} \tilde{\epsilon}_{q+1}}{\cal A}_{\rm G}(\sigma_{no}/q)&=&\partial_{k_{q} \tilde{\epsilon}_{q+1}}(k_{q} \cdot \partial_{k_{q-1}}) {\cal A}_{\rm G}(\sigma_{no}/q)\nn
&=&\partial_{k_{q-1} \tilde{\epsilon}_{q}}(\tilde{\epsilon}_{q} \cdot \partial_{\tilde{\epsilon}_{q+1}}){\cal A}_{\rm G}(\sigma_{no}/q)
\,,~~~~\label{useful identities}
\eea
This relation holds as long as an amplitude is linear on each polarization vector, and the fact that ${\cal A}_{\rm G}(\sigma_{no}/q)$ is independent of $k_q$ and $\tilde{\epsilon}_q$. More explicitly, the linearity on $\tilde{\epsilon}_b$ the effect of $\partial_{k_a \tilde{\epsilon}_b}$ is turning $k_a \cdot\tilde{\epsilon}_b$ to $1$ and annihilating all other Lorentz invariants which involve $\tilde{\epsilon}_b$. Using this, we see the validity of \eref{useful identities}.

\textbf{1}. Calculate $\tilde{\cal T}_{1}[\sigma] {\cal A}_{\rm G}^{(0)}(\sigma_{no})$ :

The expansion in \eref{Ag0 expand Aym}, together with the relation \eref{require2}, imply $\tilde{\cal T}_{1}[\sigma] {\cal A}_{\rm G}^{(0)}(\sigma_{no})=0$. Let us verify it by the following direct evaluation,
\bea
\tilde{\cal T}_{1}[\sigma] {\cal A}_{\rm G}^{(0)}(\sigma_{no})={1 \over {\tau}^{2}}\left[\partial_{k_{q-1} \tilde{\epsilon}_{q}} {(\epsilon_{q} \cdot k_{q-1})(\tilde{\epsilon}_{q} \cdot k_{q-1}) \over s_{q(q-1)}}\right] \prod \limits_{i \neq 1,q,n}\partial_{k_{i-1} \tilde{\epsilon}_{i}} \tilde{\cal T}_{1n} {\cal A}_{\rm G}(\sigma_{no}/q)=0
\,.~~~~\label{composed operator 1 1 acting on Ag0}
\eea
This result is obvious because ${\cal A}_{\rm G}^{(0)}(\sigma_{no})$ does not include $\tilde{\epsilon}_{q+1} \cdot k_{q}$.

\textbf{2}. Calculate $\tilde{\cal T}_{1}[\sigma] {\cal A}_{\rm G}^{(1)}(\sigma_{no})$ :

The expansion in \eref{Ag0 expand Aym}, and the relation \eref{require2}, also indicate that $\tilde{\cal T}_{1}[\sigma] {\cal A}_{\rm G}^{(1)}(\sigma_{no})={\cal A}_{\rm YM}^{(0)}(\sigma)$, and we now verify it as follows,
\bea
{\cal A}_{\rm YM}^{(0)}(\sigma)&=&\tilde{\cal T}_{1}[\sigma] {\cal A}_{\rm G}^{(1)}(\sigma_{no})\nn
&=&\sum_{j} (\epsilon_{q} \cdot k_{j}) \tilde{\cal T}_{1}[\sigma] \left[ {\tilde{\epsilon}_{q} \cdot \tilde{L}_{j} \cdot k_{q} \over s_{jq}}{\cal A}_{\rm G}(\sigma_{no}/q)\right]\nn
&=&\left[{\epsilon_{q} \cdot k_{q-1} \over s_{q(q-1)}}\tilde{\cal T}_{1}[\sigma](\tilde{\epsilon}_{q} \cdot \tilde{L}_{q-1} \cdot k_{q})+{\epsilon_{q} \cdot k_{q+1} \over s_{q(q+1)}}\tilde{\cal T}_{1}[\sigma](\tilde{\epsilon}_{q} \cdot \tilde{L}_{q+1} \cdot k_{q})\right]{\cal A}_{\rm G}(\sigma_{no}/q)\nn
&=&{\epsilon_{q} \cdot k_{q-1} \over s_{q(q-1)}}\tilde{\cal T}_{1}[\sigma](-k_{q-1} \cdot \tilde{f}_{q} \cdot \partial_{k_{q-1}}){\cal A}_{\rm G}(\sigma_{no}/q)\nn
& &+{\epsilon_{q} \cdot k_{q+1} \over s_{q(q+1)}}\tilde{\cal T}_{1}[\sigma](-\tilde{\epsilon}_{q+1} \cdot \tilde{f}_{q} \cdot \partial_{\tilde{\epsilon}_{q+1}}) {\cal A}_{\rm G}(\sigma_{no}/q)\nn
&=&{\epsilon_{q} \cdot k_{q-1} \over \tau s_{q(q-1)}} \prod \limits_{i \neq 1,q,q+1,n}\partial_{k_{i-1} \tilde{\epsilon}_{i}} \tilde{\cal T}_{1n} \partial_{k_{q} \tilde{\epsilon}_{q+1}}(k_{q} \cdot \partial_{k_{q-1}}){\cal A}_{\rm G}(\sigma_{no}/q)\nn
& &-{\epsilon_{q} \cdot k_{q+1} \over \tau s_{q(q+1)}}\prod \limits_{i \neq 1,q,q+1,n}\partial_{k_{i-1} \tilde{\epsilon}_{i}} \tilde{\cal T}_{1n} \partial_{k_{q-1} \tilde{\epsilon}_{q}}(\tilde{\epsilon_{q}} \cdot \partial_{\tilde{\epsilon}_{q+1}}){\cal A}_{\rm G}(\sigma_{no}/q)\nn
&=&\left({\epsilon_{q} \cdot k_{q-1} \over \tau s_{q(q-1)}}-{\epsilon_{q} \cdot k_{q+1} \over \tau s_{q(q+1)}}\right)\prod \limits_{i \neq 1,q,q+1,n}\partial_{k_{i-1} \tilde{\epsilon}_{i}} \tilde{\cal T}_{1n} \partial_{k_{q-1} \tilde{\epsilon}_{q+1}}{\cal A}_{\rm G}(\sigma_{no}/q)\nn
&=&{1 \over \tau}\left({\epsilon_{q} \cdot k_{q-1} \over s_{q(q-1)}}-{\epsilon_{q} \cdot k_{q+1} \over s_{q(q+1)}}\right) {\cal A}_{\rm YM}(\sigma/q)
\,,~~~~\label{composed operator 1 1 acting on Ag1 end}
\eea
where we use \eref{useful identities}, ${\cal T}_{1}[\sigma/q]=\prod \limits_{i \neq 1,q,q+1,n}\partial_{k_{i-1} \epsilon_{i}} {\cal T}_{1n} \partial_{k_{q-1} \epsilon_{q+1}}$ and ${\cal A}_{\rm YM}(\sigma/q)=\tilde{\cal T}_{1}[\sigma/q] {\cal A}_{\rm G}(\sigma_{no}/q)$. In second and third lines, we have dropped various terms which are annihilated by $\tilde{{\cal T}}_{1}[\sigma]$. It is exactly what we expected that \eref{composed operator 1 0 acting on Ag0}=\eref{composed operator 1 1 acting on Ag1 end}.

\textbf{3}. Calculate $\tilde{\cal T}_{1}[\sigma] {\cal A}_{\rm G}^{(2)}(\sigma_{no})$:

At this order, we need to verify $\tilde{\cal T}_{1}[\sigma] {\cal A}_{\rm G}^{(2)}(\sigma_{no})={\cal A}_{\rm YM}^{(1)}(\sigma_{no})$,
\bea
{\cal A}_{\rm YM}^{(1)}(\sigma)&=&\tilde{\cal T}_{1}[\sigma] {\cal A}_{\rm G}^{(2)}(\sigma_{no})\nn
&=&\sum_{j}{{\epsilon}_{q} \cdot L_{j} \cdot k_{q} \over s_{jq}} \tilde{\cal T}_{1}[\sigma]\Big[(\tilde{\epsilon}_{q} \cdot \tilde{L}_{j} \cdot k_{q}){\cal A}_{\rm G}(\sigma_{no}/q)\Big]\nn
&=&{{\epsilon}_{q} \cdot L_{q-1} \cdot k_{q} \over s_{q(q-1)}} \tilde{\cal T}_{1}[\sigma]\Big[(-k_{q-1} \cdot \tilde{f}_{q} \cdot \partial_{k_{q-1}}){\cal A}_{\rm G}(\sigma_{no}/q)\Big]\nn
& &+{{\epsilon}_{q} \cdot L_{q+1} \cdot k_{q} \over s_{q(q+1)}} \tilde{\cal T}_{1}[\sigma]\Big[(-\tilde{\epsilon}_{q+1} \cdot \tilde{f}_{q} \cdot \partial_{\tilde{\epsilon}_{q+1}}){\cal A}_{\rm G}(\sigma_{no}/q)\Big]\nn
&=&\left({{\epsilon}_{q} \cdot L_{q-1} \cdot k_{q} \over s_{q(q-1)}}-{{\epsilon}_{q} \cdot L_{q+1} \cdot k_{q} \over s_{q(q+1)}}\right){\cal A}_{\rm YM}(\sigma/q)\nn
&=&\left({{\epsilon}_{q} \cdot J_{q-1} \cdot k_{q} \over s_{q(q-1)}}-{{\epsilon}_{q} \cdot J_{q+1} \cdot k_{q} \over s_{q(q+1)}}\right){\cal A}_{\rm YM}(\sigma/q)
\,.~~~~\label{composed operator 1 2 acting on Ag2}
\eea
As expected, the above result is coincide with the previous one in \eref{composed operator 1 0 acting on Ag1}.

%%%%%%%%%%%%%%%%%%%%%%%%%%%%%%%%%%%%
\section{From the soft theorems of Yang-Mills to BAS}
\label{Chapter 4}
%%%%%%%%%%%%%%%%%%%%%%%%%%%%%%%%%%%%

The purpose of this section is to investigate the soft theorem of BAS amplitudes from that of YM amplitudes. We still use two combinatorial operators ${\cal T}_{0}[\sigma]$ and ${\cal T}_{1}[\sigma]$, to generate the soft behavior of BAS amplitudes along the similar line. The approach to studying the soft behavior of BAS amplitude is similar to section \ref{Chapter 3}.

As introduced in section \ref{Chapter 2, Section2}, the CHY integrand of BAS theory is $\mathrm{PT}(\sigma)\mathrm{PT}(\sigma')$. It means each BAS amplitude carries two orderings, expressed as ${\cal A}_{\rm BAS}(\sigma|\sigma')$. Without losing of generality, we fix one of them as $\sigma=\{1,\cdots,n\}$, and leave another one to be arbitrary ordering $\sigma'$.

Using the expansion in \eref{expand G and YM}, we expect that
\bea
{\cal A}_{\rm YM}^{(0)}(\sigma')&=&\sum_{\sigma'}C^{(0)}(\epsilon,k;\sigma'){\cal A}^{(0)}_{\rm BAS}(\sigma|\sigma')\,,\nn
{\cal A}_{\rm YM}^{(1)}(\sigma')&=&\sum_{\sigma'}C^{(0)}(\epsilon,k;\sigma'){\cal A}^{(1)}_{\rm BAS}(\sigma|\sigma')
+\sum_{\sigma'}C^{(1)}(\epsilon,k;\sigma'){\cal A}^{(0)}_{\rm BAS}(\sigma|\sigma')\,.
\eea
Then, the argument similar to that in section \ref{subsec-e-We} leads to
\bea
{\cal T}_0[\sigma]{\cal A}_{\rm YM}^{(0)}(\sigma')&=&{\cal A}_{\rm BAS}^{(0)}(\sigma|\sigma')\,,\nn
{\cal T}_0[\sigma]{\cal A}_{\rm YM}^{(1)}(\sigma')&=&{\cal A}_{\rm BAS}^{(1)}(\sigma|\sigma')\,,
\eea
and
\bea
{\cal T}_1[\sigma]{\cal A}_{\rm YM}^{(1)}(\sigma')&=&{\cal A}_{\rm BAS}^{(0)}(\sigma|\sigma')\,.
\eea
%

%%%%%%%%%%%%%%%%%%%%%%%%%%%%%%%%%%%%
\subsection{From YM to BAS: operator ${\cal T}_0[\sigma]$}
\label{Chapter 4, Section 1}
%%%%%%%%%%%%%%%%%%%%%%%%%%%%%%%%%%%%

Now we use the combinatorial operator ${\cal T}_{0}[\sigma]$ to generate ${\cal A}_{\rm BAS}^{(0)}(\sigma|\sigma')$ and ${\cal A}_{\rm BAS}^{(1)}(\sigma|\sigma')$ from ${\cal A}_{\rm YM}^{(0)}(\sigma')$ and ${\cal A}_{\rm YM}^{(1)}(\sigma')$.

\textbf{1}. Calculate ${\cal T}_{0}[\sigma] {\cal A}_{\rm YM}^{(0)}(\sigma')$ :
\bea
{\cal A}_{\rm BAS}^{(0)}(\sigma|\sigma')&=&{\cal T}_{0}[\sigma] {\cal A}_{\rm YM}^{(0)}(\sigma')\nn
&=&{1 \over \tau} \left({\cal T}_{(q-1)q(q+1)} \sum_{j} {\epsilon_{q} \cdot k_{j} \over s_{jq}} \Delta_{jq}\right) {\cal T}[\sigma/q] {\cal A}_{\rm YM}(\sigma'/q)\nn
&=&{1 \over \tau} \left({\Delta_{(q-1)q} \over s_{q(q-1)}}+{\Delta_{q(q+1)} \over s_{q(q+1)}}\right) {\cal A}_{\rm BAS}(\sigma/q|\sigma'/q)
\,.~~~~\label{composed operator 1 0 acting on Aym0}
\eea
The result of \eref{composed operator 1 0 acting on Aym0} is the same with that in \cite{Zhou:2022orv}. Note that $\Delta_{jq}$ depends on the ordering $\sigma'$. This result indicates that the BAS amplitude satisfies the soft theorem at the leading order (proportional to $\tau^{-1}$). 

\textbf{2}. Calculate ${\cal T}_{0}[\sigma] {\cal A}_{\rm YM}^{(1)}(\sigma')$ :
\bea
{\cal A}_{\rm BAS}^{(1)}(\sigma|\sigma')&=&{\cal T}_{0}[\sigma]{\cal A}_{\rm YM}^{(1)}(\sigma')\nn
&=&\left({\cal T}_{(q-1)q(q+1)} \sum_{j} {\epsilon_{q} \cdot J_{j} \cdot k_{q} \over s_{jq}} \Delta_{jq}\right){\cal T}[\sigma/q]{\cal A}_{\rm YM}(\sigma'/q)\nn
&=&\left({\cal T}_{(q-1)q(q+1)} \sum_{j} {-k_{j} \cdot f_{q} \cdot \partial_{k_{j}} \over s_{jq}} \Delta_{jq}\right){\cal A}_{\rm BAS}(\sigma/q|\sigma'/q)\nn
&=& \left\{{\Delta_{(q-1)q} \over s_{q(q-1)}}(k_{q} \cdot \partial_{k_{q-1}})+{\Delta_{q(q+1)} \over s_{q(q+1)}}(k_{q} \cdot \partial_{k_{q+1}}) \right.\nn
& &-\left.{1 \over 2} {\cal T}_{(q-1)q(q+1)}\left[ \left(\epsilon_{q} \cdot \partial_{k_{q-1}} \right) \Delta_{(q-1)q}+\left(\epsilon_{q} \cdot \partial_{k_{q+1}} \right) \Delta_{(q+1)q} \right] \right\} {\cal A}_{\rm BAS}(\sigma/q|\sigma'/q)\nn
&\overset{\text{?}}{=}&S^{(1)}_{s}(q){\cal A}_{\rm BAS}(\sigma/q|\sigma'/q)
\,,~~~~\label{composed operator 1 0 acting on Aym1}
\eea
where $\overset{\text{?}}{=}$ means it is not very appropriate to regard $S^{(1)}_{s}(q)$ as an universal soft factor, since the $4$-point BAS amplitudes have vanishing soft behavior at the sub-leading order. Indeed, the corresponding CHY integrand $\mathrm{PT}(\sigma)\mathrm{PT}(\sigma')$ ensures that the BAS amplitudes only feature massless propagators arise from scattering equations, thus for the $4$-point soft behaviors, the associated $3$-point sub-amplitudes ${\cal A}_{\rm BAS}(\sigma/q|\sigma'/q)$ are constants, which are annihilated by differential operators in \eref{composed operator 1 0 acting on Aym1}. On the other hand, we find that for $n\geq5$, the sub-leading soft behavior of BAS amplitudes
always satisfy the representation in \eref{composed operator 1 0 acting on Aym1}. We have verified the above sub-leading soft behavior in \eref{composed operator 1 0 acting on Aym1} until $8$-point, by using Feynman diagrams, according to the definition of BAS amplitudes.

Thus we conclude the leading soft behavior of BAS amplitudes factorize as
\bea
{\cal A}^{(0)}_{\rm BAS}(\sigma|\sigma')=S^{(0)}_s(q){\cal A}_{\rm BAS}(\sigma/q|\sigma'/q)\,,
\eea
with the universal soft factor
\bea
{S}_{s}^{(0)}(q)={1 \over \tau}\left({\Delta_{(q-1)q} \over s_{q(q-1)}}+{\Delta_{q(q+1)} \over s_{q(q+1)}}\right)
\,.~~~~\label{leading-order soft factor of Abas}
\eea
%

%%%%%%%%%%%%%%%%%%%%%%%%%%%%%%%%%%%%
\subsection{From YM to BAS: operator ${\cal T}_1[\sigma]$}
\label{Chapter 4, Section 2}
%%%%%%%%%%%%%%%%%%%%%%%%%%%%%%%%%%%%

Then we turn to the operator ${\cal T}_1[\sigma]$.

\textbf{1}. Calculate ${\cal T}_{1}[\sigma] {\cal A}_{\rm YM}^{(0)}(\sigma')$ :

The expectation ${\cal T}_{1}[\sigma] {\cal A}_{\rm YM}^{(0)}(\sigma')$ can be directly verified as,
\bea
{\cal T}_{1}[\sigma] {\cal A}_{\rm YM}^{(0)}(\sigma')={1 \over {\tau}^{2}} \left(\partial_{k_{q-1}\epsilon_{q}} \sum_{j} {\epsilon_{q} \cdot k_{j} \over s_{jq}} \Delta_{jq}\right) \prod \limits_{i \neq 1,q,n}\partial_{k_{i-1} {\epsilon}_{i}} {\cal T}_{1n} {\cal A}_{\rm YM}(\sigma'/q)=0
\,.~~~~\label{composed operator 1 1 acting on Aym0}
\eea

\textbf{2}. Calculate ${\cal T}_{1}[\sigma] {\cal A}_{\rm YM}^{(1)}(\sigma')$ :

Then we verify that that action ${\cal T}_{1}[\sigma] {\cal A}_{\rm YM}^{(1)}(\sigma')$ gives arise to the soft behavior in $\eref{composed operator 1 0 acting on Aym0}$.
\bea
{\cal A}_{\rm BAS}^{(0)}(\sigma|\sigma')&=&{\cal T}_{1}[\sigma] {\cal A}_{\rm YM}^{(1)}(\sigma')\nn
&=&{1 \over \tau} \prod \limits_{i \neq 1,q,q+1,n}\partial_{k_{i-1} {\epsilon}_{i}} {\cal T}_{1n} (\partial_{k_{q-1} {\epsilon}_{q}} \partial_{k_{q} {\epsilon}_{q+1}}) \left[{(k_{q-1} \cdot \epsilon_{q})(k_{q} \cdot \partial_{k_{q-1}}) \over s_{q(q-1)}} \Delta_{(q-1)q}\right.\nn
& &\left.+{(\epsilon_{q+1} \cdot k_{q})(\epsilon_{q} \cdot \partial_{\epsilon_{q+1}}) \over s_{q(q+1)}} \Delta_{q(q+1)}\right] {\cal A}_{\rm YM}(\sigma'/q)\nn
&=&{1 \over \tau} \prod \limits_{i \neq 1,q,q+1,n}\partial_{k_{i-1} {\epsilon}_{i}} {\cal T}_{1n} \partial_{k_{q-1} {\epsilon}_{q+1}}\left({\Delta_{(q-1)q} \over s_{q(q-1)}}+{\Delta_{q(q+1)} \over s_{qb}}\right) {\cal A}_{\rm YM}(\sigma'/q)\nn
&=&{1 \over \tau} \left({\Delta_{(q-1)q} \over s_{q(q-1)}}+{\Delta_{q(q+1)} \over s_{q(q+1)}}\right) {\cal T}_{1}[\sigma/q] {\cal A}_{\rm YM}(\sigma'/q)\nn
&=&{1 \over \tau} \left({\Delta_{(q-1)q} \over s_{q(q-1)}}+{\Delta_{q(q+1)} \over s_{q(q+1)}}\right) {\cal A}_{\rm BAS}(\sigma/q|\sigma'/q)
\,.~~~~\label{composed operator 1 1 acting on Aym1}
\eea
As expected, the above result is coincide with the previous one in \eref{composed operator 1 0 acting on Aym0}.

%%%%%%%%%%%%%%%%%%%%%%%%%%%%%
\section{Summary}
\label{Chapter 6}
%%%%%%%%%%%%%%%%%%%%%%%%%%%%

Starting from three soft theorems of GR amplitudes, we used transmutation operators to investigate soft behaviors of YM and BAS amplitudes. The resulted soft factors are coincide with those in literatures, and also serve as the verification of theoretical self-consistency. Beyond this, we find that the YM amplitudes do not have universal soft behavior at the sub-sub-leading order, while BAS amplitudes do not have universal soft behaviors at the sub-leading order, due to the speciality of $4$-point amplitudes. Comparing with our recent work \cite{Wei:2024ynm}, the treatment in this note explicitly tells us the universality of soft behavior at higher order is broken by what specific reason for the $4$-point case. An interesting observation is that for amplitudes which include at least $n\geq5$ external scalars, each BAS amplitude can be represented as an universal operator act on the $(n-1)$-point amplitude, at the sub-leading order.

%It is natural to ask whether YM amplitudes also have uniformed soft behavior at higher order, for $n\geq I$, where $I$ is an integer satisfying $I\geq5$. Along the line of this note, the sub-sub-leading YM soft behavior can be generated by acting $\tilde{\cal T}_0[\sigma]$ on ${\cal A}_{\rm G}^{(2)}(\sigma_{no})$. The difficulty is that how to express ${\cal A}_{\rm G}^{(2)}(\sigma_{no})$ in the form with distinguished $\epsilon$ and $\tilde{\epsilon}$. However, we have another alternative way to study ${\cal A}_{\rm YM}^{(2)}(\sigma)$. Since the operator ${\cal T}_1[\sigma]$ links ${\cal A}_{\rm YM}^{(2)}(\sigma)$ and ${\cal A}_{\rm BAS}^{(1)}(\sigma|\sigma')$ together, we can solve ${\cal A}_{\rm YM}^{(2)}(\sigma)$ from \eref{composed operator 1 0 acting on Aym1} (for $n\geq5$). We leave this issue to the future work.

An interesting potential application of soft behavior in \eref{composed operator 1 0 acting on Aym1} is the construction for $1$-loop integrand of Feynman integration. It has been widely studied that the tree amplitudes of various theories can be constructed by reversing soft theorems \cite{Zhou:2022orv,Zhou:2023quv}. A natural generalization should be, wether the similar idea can be used to calculate integrands at loop level. Suppose we use the well known forward limit method to generate $1$-loop BAS integrands from tree amplitudes, then one can expect that the sub-leading soft behavior in \eref{composed operator 1 0 acting on Aym1} will play the important role. The reason is, when the soft leg is connected to loop propagators, it does not cause any divergence in the soft limit, thus the corresponding soft behavior contributes to the sub-leading order rather than the leading one. On the other hand, the disadvantage that \eref{composed operator 1 0 acting on Aym1} is not satisfied by the $4$-point case can be ignored, since the $4$-point tree amplitudes correspond to $1$-loop bubbles in the forward limit. Such $1$-loop bubbles are called the scale less integrals, which are integrated to zero under the dimensional regularization, thus are un-physical \cite{Leibbrandt:1975dj}.

\appendix
%%%%%%%%%%%%%%%%%%%%%%%%%%%%%
\section*{Appendix}
%%%%%%%%%%%%%%%%%%%%%%%%%%%%
As a further verification, we consider another combinatorial operator,
\bea
{\cal T}_{2}[\sigma]&=&{\cal T}_{1n} \prod \limits_{i \neq 1,q-1,q,q+1,n} \partial_{k_{i-1} \epsilon_{i}} {\cal T}_{(q-2)(q-1)q} {\cal T}_{(q-2)q(q+1)} \partial_{k_{q-2} \epsilon_{q+1}}\nn
&=&{\cal T}_{1n} \prod \limits_{i \neq 1,q-1,q,q+1,n} \partial_{k_{i-1} \epsilon_{i}} {\cal T}_{(q-2)q(q+1)} \partial_{k_{q-2} \epsilon_{q+1}} \partial_{k_{q-2} \epsilon_{q-1}}\nn
& &-{\cal T}_{1n} \prod \limits_{i \neq 1,q-1,q,q+1,n} \partial_{k_{i-1} \epsilon_{i}} {\cal T}_{(q-2)q(q+1)} \partial_{k_{q-2} \epsilon_{q+1}} \partial_{k_{q} \epsilon_{q-1}}\nn
&=&{\cal T}_{20}[\sigma]+{\cal T}_{21}[\sigma]
\,,~~~~\label{composed operator 2}
\eea
with
\bea
{\cal T}_{20}[\sigma]={\cal T}_{1n} \prod \limits_{i \neq 1,q-1,q,q+1,n} \partial_{k_{i-1} \epsilon_{i}} {\cal T}_{(q-2)q(q+1)} \partial_{k_{q-2} \epsilon_{q+1}} \partial_{k_{q-2} \epsilon_{q-1}}
\,,~~~~\label{composed operator 2 0}
\eea
and
\bea
{\cal T}_{21}[\sigma]=-{\cal T}_{1n} \prod \limits_{i \neq 1,q-1,q,q+1,n} \partial_{k_{i-1} \epsilon_{i}} {\cal T}_{(q-2)q(q+1)} \partial_{k_{q-2} \epsilon_{q+1}} \partial_{k_{q} \epsilon_{q-1}}
\,,~~~~\label{composed operator 2 1}
\eea
The full operator ${\cal T}_{2}[\sigma]$ is decomposed into two parts, one is ${\cal T}_{20}[\sigma] \propto \tau^{0}$, and another one is ${\cal T}_{21}[\sigma] \propto \tau^{-1}$. After a few algebra, we find,
\bea
& &{\cal T}_{20}[\sigma]{\cal A}_{\rm YM}^{(0)}(\sigma')+{\cal T}_{21}[\sigma]{\cal A}_{\rm YM}^{(1)}(\sigma')\nn
&=&-{\Delta_{(q+1)q} \over \tau s_{q(q+1)}}{\cal T}_{1n} \prod \limits_{i \neq 1,q-1,q,q+1,n} \partial_{k_{i-1} \epsilon_{i}} \partial_{k_{q-2} \epsilon_{q-1}}\partial_{k_{q-2} \epsilon_{q+1}} {\cal A}_{\rm YM}(\sigma'/q)\nn
& &+{\Delta_{(q-1)q} \over \tau s_{q(q-1)}}{\cal T}_{1n} \prod \limits_{i \neq 1,q-1,q,q+1,n} \partial_{k_{i-1} \epsilon_{i}} \partial_{k_{q-2} \epsilon_{q-1}}\partial_{k_{q-2} \epsilon_{q+1}} {\cal A}_{\rm YM}(\sigma'/q)\nn
& &-{\Delta_{(q-1)q} \over \tau s_{q(q-1)}}{\cal T}_{1n} \prod \limits_{i \neq 1,q-1,q,q+1,n} \partial_{k_{i-1} \epsilon_{i}} \partial_{k_{q+1} \epsilon_{q-1}}\partial_{k_{q-2} \epsilon_{q+1}} {\cal A}_{\rm YM}(\sigma'/q)\nn
& &+{\Delta_{(q+1)q} \over \tau s_{q(q+1)}}{\cal T}_{1n} \prod \limits_{i \neq 1,q-1,q,q+1,n} \partial_{k_{i-1} \epsilon_{i}} \partial_{k_{q+1} \epsilon_{q-1}}\partial_{k_{q-2} \epsilon_{q+1}} {\cal A}_{\rm YM}(\sigma'/q)\nn
&=&{1 \over \tau}\left({\Delta_{(q-1)q} \over s_{q(q-1)}}+{\Delta_{q(q+1)} \over s_{q(q+1)}}\right){\cal T}_{1n} \prod \limits_{i \neq 1,q-1,q,q+1,n} \partial_{k_{i-1} \epsilon_{i}} {\cal T}_{(q-2)(q-1)(q+1)}\partial_{k_{q-2} \epsilon_{q+1}} {\cal A}_{\rm YM}(\sigma'/q)\nn  
&=&S^{(0)}_{s}(q) {\cal T}_{0}[\sigma/q] {\cal A}_{\rm YM}(\sigma'/q)\nn
&=&{\cal A}_{\rm BAS}^{(0)}(\sigma|\sigma')
\,,~~~\label{T2 and Abas0}
\eea
where we used \eref{leading-order soft factor of Aym}, \eref{sub-leading-order soft factor of Aym}, \eref{leading-order soft factor of Abas} and ${\cal T}_{0}[\sigma/q]={\cal T}_{1n} \prod \limits_{i \neq 1,q-1,q,q+1,n} \partial_{k_{i-1} \epsilon_{i}} {\cal T}_{(q-2)(q-1)(q+1)}\partial_{k_{q-2} \epsilon_{q+1}}$ (pay attention to different for ${\cal T}_{0}[\sigma/q]$ and ${\cal T}[\sigma/q]$), as well as ${\cal T}_{0}[\sigma/q] {\cal A}_{\rm YM}(\sigma'/q)={\cal A}_{\rm BAS}(\sigma/q|\sigma'/q)$. It is satisfy our expectation based on discussions in section \ref{subsec-e-We}. Similarly, we have
\bea
\tilde{{\cal T}}_{20}[\sigma]{\cal A}_{\rm G}^{(0)}(\sigma_{no})+\tilde{{\cal T}}_{21}[\sigma]{\cal A}_{\rm G}^{(1)}(\sigma_{no})={\cal A}_{\rm YM}^{(0)}(\sigma)
\,,~~~\label{T2 and Aym0}
\eea
and
\bea
\tilde{{\cal T}}_{20}[\sigma]{\cal A}_{\rm G}^{(1)}(\sigma_{no})+\tilde{{\cal T}}_{21}[\sigma]{\cal A}_{\rm G}^{(2)}(\sigma_{no})={\cal A}_{\rm YM}^{(1)}(\sigma)
\,.~~~\label{T2 and Aym1}
\eea
The calculation process is similar to \eref{T2 and Abas0}, just need to do $\Delta_{jq} \to \epsilon_{q} \cdot k_{j}$ and $\Delta_{jq} \to \epsilon_{q} \cdot L_{j} \cdot k_{q}$. The term $j=q-2$ will be counteracted out, leaving only terms $j=q-1$ and $j=q+1$.

There is a clear difference between ${\cal T}_{2}[\sigma]$  and ${\cal T}_{0}[\sigma]$, ${\cal T}_{1}[\sigma]$. the operator ${\cal T}_{2}[\sigma]$ acts on at least $5$-point amplitudes, because it involves at least 3 insertion operators. The combinatorial operator proportional to $\tau^{-2}$ does not exist, however, one can always constructed the combinatorial operator which can be split into two parts, one is proportional to $\tau^{-2}$ and another one is proportional to $\tau^{-1}$. Just like the above ${\cal T}_{2}[\sigma]$, these operators make sense for $n$-point amplitudes with $n\geq5$.

\end{document}